\documentclass[conference, onecolumn, draft]{IEEEtran}
\IEEEoverridecommandlockouts 
\usepackage{cite}
\usepackage{graphicx}
\usepackage{url}
\usepackage{cite}
\usepackage{graphicx}
\usepackage{amsmath}
\usepackage{amssymb}
\usepackage{subfigure}
\usepackage{lipsum, color}
\usepackage{mathtools, cuted}
\usepackage{textcase}
\usepackage{algorithm}
\usepackage{enumerate}
\usepackage{balance}

\usepackage{dsfont}
\usepackage{algpseudocode}
\def \ba           {\boldsymbol{a}}

\def \bd           {\boldsymbol{d}}

\def \bs           {\boldsymbol{s}}

\def \bu           {\boldsymbol{u}}

\def \bw           {\boldsymbol{w}}

\def \by           {\boldsymbol{y}}

\def \bA           {\boldsymbol{A}}

\def \bD           {\boldsymbol{D}}

\def \bI           {\boldsymbol{I}}
\def \bJ           {\boldsymbol{J}}

\def \bQ           {\boldsymbol{Q}}
\def \bR           {\boldsymbol{R}}
\def \bS           {\boldsymbol{S}}
\def \bT           {\boldsymbol{T}}

\def \bW           {\boldsymbol{W}}

\def \balpha       {\boldsymbol{\alpha}}

\def \bGamma       {\boldsymbol{\Gamma}}

\def \bepsilon     {\boldsymbol{\epsilon}}

\def \bdeta        {\boldsymbol{\eta}}

\def \bSigma       {\boldsymbol{\Sigma}}

\def \bupsilon     {\boldsymbol{\upsilon}}

\def \bchi         {\boldsymbol{\chi}}

\def \bzero        {\boldsymbol{0}}

\def \complexC     {\mathds{C}}
\def \expecE       {\mathds{E}}

\def \tr           {\mathrm{tr}}

\def \diag         {\mathrm{diag}}
\def \Diag         {\mathrm{Diag}}

\def \asin         {\sin^{-1}}

\def \half         {\frac{1}{2}}
\def \roothalf     {\frac{1}{\sqrt{2}}}

\def \st           {\mathrm{s.t.}}

\def \sign         {\mathrm{sign}}

\newcommand\undermat[2]{
	\makebox[0pt][l]{${\underbrace{\hphantom{
					\begin{matrix}#2\end{matrix}}}_{\text{$#1$}}}$}#2}
\begin{document}

\title{\huge Waveform Design for One-Bit Radar Systems Under Uncertain Interference Statistics
}
\author{
\IEEEauthorblockN{Arindam~Bose\IEEEauthorrefmark{1},
Aria~Ameri\IEEEauthorrefmark{2}, and Mojtaba~Soltanalian\IEEEauthorrefmark{3}}
\thanks{%
	\IEEEauthorrefmark{1}Corresponding author (e-mail: \textit{abose4@uic.edu}).
}  
\thanks{%
	This work was supported in part by U.S. National Science Foundation Grants CCF-1704401 and ECCS-1809225.
}
\IEEEauthorblockA{
\IEEEauthorrefmark{0}Department of Electrical and Computer Engineering, University of Illinois at Chicago, Chicago, IL 60607, USA
\\ Email: \{\IEEEauthorrefmark{1}abose4, \IEEEauthorrefmark{2}aameri2, \IEEEauthorrefmark{3}msol\}@uic.edu}}

\maketitle

\begin{abstract}
An important problem in cognitive radar is to enhance the estimation performance of the system by a joint design of its probing signal and receive filter using the a priori information on interference.
In such cases, the knowledge of interference statistics (particularly the covariance) plays a vital role in an effective design of the radar waveforms.
In most practical scenarios, however, the received signal and interference statistics are available subject to some uncertainty.
An extreme manifestation of this practical observation occurs for radars employing one-bit receivers, where only a normalized version of interference covariance matrix can be obtained.
In this paper, we formulate a waveform optimization problem and devise an algorithm to design the transmit waveform and the receive filter of one-bit radars given such uncertainties in acquired interference statistics.
The effectiveness of the proposed algorithm is corroborated through numerical analysis.
\end{abstract}

\begin{IEEEkeywords}
	Cognitive radar, clutter rejection, joint design, probing signal, receive filter
\end{IEEEkeywords}

\section{Introduction and Prior Works}
In cognitive active sensing applications, an important problem is to jointly design the probing sequence and the receive filter using the \textit{apriori} knowledge of clutter and interference in order to minimize the estimation error of the target parameters \cite{6104176, 4644058, 6472022}.
Clutter refers to the unwanted echoes that are usually correlated with the transmitted waveform, while the signal independent noise as well as (adverse) jamming signals are termed as interference \cite{4644058}.
A natural way to minimize the effects of clutter and interference is to maximize the signal-to-clutter-plus-interference ratio (SCIR) of the receiver output.
It is well known that a matched filter (MF) can maximize the signal-to-noise (SNR) in the presence of additive white noise, it however, fails to perform well in the case of clutter or jamming suppression.
As an alternative, one can use a \textit{mismatched} filter (MMF) at the receiver by trading off SNR for SCIR \cite{4644058}.
In comparison to MF, an MMF allows more degrees of freedom by introducing a receive filter and is not subject to various power constraints of the transmit waveform such as constant-modulus or low peak-to-average ratio (PAR) constraint.
Thus, a joint design of the transmit waveform and the MMF receive filter can offer a more efficient parameter estimation framework \cite{Spafford}.
%

In \cite{6104176}, the authors presented a joint design scheme of the receive filter and transmit waveform by minimizing the mean-square error (MSE) of the estimate of a target's scattering coefficient in the presence of clutter and interference subject to some practical constraints such as constant-modulus or low PAR constraint on the transmit signal. 
To this end, they presented three flavors of their algorithm: Cognitive REceiver and Waveform design (CREW); namely, CREW (gra), CREW (fre), and CREW (mat). 
Another variation of CREW; namely, CREW (cyclic) can be found in \cite{6472022}, where the authors formulated a cyclic approach to jointly design the transmit waveform and receive filter coefficients.
Note that in all the aforementioned techniques, the receiver is assumed to have high precision analog-to-digital converters (ADC). In other words the quantization noise is modeled as additive noise that usually has little to no impact on algorithms that assume the infinite precision case, provided that the sampling resolution is high enough.
The assumption of high-precision data is, however, inappropriate when the measurements are extremely quantized to very low bit-rates.
In the most extreme case, the sampling process is done by utilizing a simple sign comparator and the received signal is represented using only one bit per sample \cite{8822763, 8683876}.
One-bit quantizers on one hand, are not only low-cost and low-power hardware components, but also much faster than traditional scalar quantizers, resulting in great reduction in the complexity of hardware implementation.
On the other hand, it is now well known that signals can be recovered with high accuracy from one-bit measurements, at a slightly increased computational cost \cite{5955138}.
This increased cost incurs from the fact that by using a one-bit receiver, the knowledge of interference statistics are available in only a \textit{normalized} sense and such uncertainties prohibit one from using traditional algorithms.\nocite{2019arXiv191007591B, 8645383}

In the subsequent, we propose a specialized variation of CREW (cyclic) \cite{6472022} to tackle the problem of jointly designing the probing signals and the receive filter coefficients in the presence of uncertainty in interference statistics.

\textit{Notation:} We use bold-lowercase and bold-uppercase letters to represent vectors and matrices, respectively. The superscripts $(\cdot)^*$, $(\cdot)^T$, and $(\cdot)^H$ represent the conjugate, the transpose, and the Hermitian operator. $\sign(\cdot), \Re(\cdot)$, and $\Im(\cdot)$ are the element-wise sign, real part and imaginary part of a complex element, respectively. $\expecE\{\cdot\}$ represents expected value of a  random variable. $\tr(\cdot)$ is the trace of a matrix. In addition, $\diag(\cdot)$ and $\Diag(\cdot)$ represent the diagonal vector of its argument matrix and the diagonal matrix made with its argument vector, respectively. $\bI$ is the identity matrix. $\complexC^N$ is the set of complex vectors of length $N$. Finally, $\odot$ represents the elementwise product.

\section{Signal Model and Problem Formulation}
Let $\bs = [s_1~s_2~ \cdots s_N]^T \in \complexC^N$ denote the transmit sequence of length $ N $, that is to be used to modulate the train of subpulses.
We adopt the discrete data model described in \cite{6104176}  in order to layout the problem formulation.
Under the assumptions of negligible intrapulse Doppler shift, and that the sampling is synchronized to the pulse rate, the received discrete-time baseband signal after pulse compression and alignment with the current range cell of interest, satisfies \begin{align}
\by = \bA^H\balpha + \bepsilon,
\end{align}
where
\begin{align}
\bA^H &= \begin{bmatrix}
	s_1 & 0 & \cdots & 0 & s_N & s_{N-1} & \cdots & s_2 \\
	s_2 & s_1 &  & \vdots & 0 & s_N &  & \vdots \\
	\vdots & \vdots & \ddots & 0 & \vdots & \vdots & \ddots & s_N \\
	s_N & s_{N-1} & \cdots & s_1 & 0 & 0 & \cdots & 0
\end{bmatrix}, \\
\balpha &= [\alpha_0~\alpha_1~\cdots~\alpha_{N-1}~\alpha_{-N+1}~\cdots~\alpha_{-1}]^T \in \complexC^{2N-1},
\end{align}
where the parameter $\alpha_0$ is the scattering coefficient of the current range cell, while $\{\alpha_k\}_{k\neq0}$ are that of the adjacent range cells contributing to the clutter, and $\bepsilon$ is the signal independent interference comprising of measurement noise as well as other disturbances such as jamming. 
In addition, we assume that $ \bGamma \triangleq \expecE\{\bepsilon\bepsilon^H\} $, and
$ \beta \triangleq \expecE\{|\alpha_k|^2\} $ for $ k\neq 0 $, and that $\bepsilon$ and $\{\alpha_k\}$ are zero-mean i.i.d.
Note that in a traditional radar system, $\beta$ and $\bGamma$ can be obtained via some prescanning procedure \cite{4644058}. 

For a known $\beta$ and $\bGamma$, the estimation of the scattering coefficient of the current range cell, $\alpha_0$, can be efficiently achieved by using an MMF, and is given as \cite{6472022},
\begin{align*}
	\hat{\alpha_0} = \frac{\bw^H\by}{\bw^H\bs},
\end{align*} 
where $\bw\in\complexC^N$ is the MMF coefficient vector.
Therefore, the MSE of estimation of $\alpha_0$ can be derived as
\begin{align}\label{eq:mse}
	\text{MSE}(\hat{\alpha_0}) = \expecE\left\lbrace \left|\frac{\bw^H\by}{\bw^H\bs} - \alpha_0\right|^2\right\rbrace = \frac{\bw^H\bR\bw}{|\bw^H\bs|^2},
\end{align}
where
\begin{align}\label{eq:R}
	\bR = \beta\sum_{\substack{k = -N+1 \\ k\neq 0}}^{N-1} {\bJ_k\bs\bs^H\bJ_k^H} + \bGamma,
\end{align} 
and $\{\bJ_k\}$ are the shift matrices satisfying,
\begin{align}
	\bJ_k=\bJ_{-k}^H=
	\begin{bmatrix}
	0 			& \dots & 0 	& 1 	& \dots 	& 0   	\\
	\vdots		& 		&	 	&	  	& \ddots 	& 		\\
	& 		&    	&	 	&			& 1  	\\
	\undermat{k}{0 & \dots 	& 0} 	& \dots	 	&	& 	 	\\
	\end{bmatrix}^H_{N \times N}, \\
	k = 0, 1, \cdots, N-1. \nonumber
	\label{eq:J_k}	
\end{align}
Note that the denominator of the MSE in \eqref{eq:mse} is the power of the signal at the receiver and its numerator is the power of the
interferences.
Therefore, minimizing the MSE is identical to maximizing the SCIR.

\subsection{One-bit receiver}
In the case of receivers with one-bit ADC, the quantizer is nothing but a simple sign comparator and each measurement is represented using only one bit, i.e., $+1$ or $-1$, and thus, the auto-correlation of the received signal is only obtainable in a \textit{normalized} sense, as described in the subsequent \cite{8822763}.

Let $Y(t)$ denote a real-valued, scalar, and stationary Gaussian process that undergoes a one-bit sampling process $Z(t)=\sign(Y(t))$.
The auto-correlation function of the process $Z(t)$, denoted by $R_Z(\tau)$, is given by
\begin{align}
R_Z(\tau) = \mathbb{E} \{ Z(t+\tau) Z(t) \} = \frac{2}{\pi}\asin{\bar{R}_Y(\tau)},
\end{align}
where $\bar{R}_Y(\tau) = R_Y(\tau) / R_Y(0)$ is the normalized auto-correlation function of the process $Y(t)$ \cite{van1966spectrum}.
On the other hand, the Bussgang theorem \cite{bussgang1952crosscorrelation} states that the cross-correlation function of the processes $Y(t)$ and $Z(t)$ is proportional to the auto-correlation function of $Y(t)$, i.e., $R_{ZY}(\tau) = \zeta R_Y(\tau)$, where the factor $\zeta$ depends on the power of the process $Y(t)$. 

\par
The case of complex-valued vector processes can be elaborated in a similar manner \cite{liu2017one}.
Let $\bupsilon$ be the one-bit sampled data obtained from $\by$ using complex one-bit ADCs at the receiver, given by
\begin{align}\label{eq:gamma}
\bupsilon = \roothalf \mathrm{csign}(\by) \triangleq \roothalf [\sign(\Re(\by)) + j\sign(\Im(\by))].
\end{align}
Let $ \bR_{\by}$ and $\bR_{\bupsilon}$ denote the auto-correlation of the processes $\by$ and $\bupsilon$, respectively. It has been shown in \cite{liu2017one} that the following equality holds:
\begin{align}
\bar{\bR}_{\by}  =  \sin \left( \frac{\pi}{2} \bR_{\bupsilon} \right),
\label{eq:RBussgang}
\end{align}
where the normalized auto-correlation matrix of $\by$ is given as
\begin{align}
	\bar{\bR}_{\by} \triangleq \bW^{-\half} \bR_{\by} \bW^{-\half},
\end{align}
and $\bW = \bR_{\by} \odot \bI$.

In the light of above, it can be verified that in the scenario of having complex one-bit sampled data, the matrix $\bR$ in \eqref{eq:mse} is obtainable only in a \textit{normalized} sense, i.e., one only has access to 
\begin{align}
	\bar{\bR} = \bD^{-\half} \bR \bD^{-\half},
	\label{eq:Rbar}
\end{align} where $\bD = \bR \odot \bI$.
Then, the problem of interest is to design the transmit waveform $\bs$ and the receive filter $\bw$ given the normalized interference statistics $\bar{\bR}$. 
In the following, we denote $\bd = \diag(\bD^{\half})$.
 
In such a case, a meaningful approach to the aforementioned design problem is to consider:
\begin{align}\label{eq:MSEonebit}
	\min_{\bw, \bs}\qquad\expecE\left\lbrace\frac{\bw^H \bD^{\half} \bar{\bR} \bD^{\half} \bw}{|\bw^H \bs|^2}\right\rbrace,
\end{align}
under some practical signal power constraint. Note that the expectation is taken over $\bD$. The above problem is clearly non-convex. In the following, we handle the non-convexity of the optimization objective in \eqref{eq:MSEonebit} with respect to (w.r.t.) the probing sequence $\bs$ and the receive filter $\bw$ using an alternating approach and propose a specialized flavor of CREW (cyclic), named as CREW (one-bit).

\section{Proposed Method: CREW (One-Bit)}

\subsection{Optimization of $\bs$}
Following \eqref{eq:R}, the numerator of \eqref{eq:mse} can be rearranged, for a fixed $\bw$, as
\begin{align}\label{eq:chi}
\bw^H\bR\bw &= \bw^H\left(\beta\sum_{\substack{k = -N+1 \\ k\neq 0}}^{N-1} {\bJ_k\bs\bs^H\bJ_k^H + \bGamma}\right)\bw\\
&= \bs^H\left(\beta\underbrace{\sum_{\substack{k = -N+1 \\ k\neq 0}}^{N-1} {\bJ_k\bw\bw^H\bJ_k^H}}_{\bchi}\right)\bs + \bw^H\bGamma\bw\nonumber.
\end{align}
Thus, the criterion in \eqref{eq:mse} can be reformulated as,
\begin{align}\label{eq:mse-aug}
\dfrac{\text{MSE}(\hat{\alpha_0})}{\beta}  = \frac{\bs^H\bchi\bs + \mu} {\bs^H\bW\bs},
\end{align}
where $\mu = (\bw^H\bGamma\bw)/\beta$ and $\bW = \bw\bw^H$.
It is interesting to note that $\mu$ is unknown; however, independent of $\bs$, and thus merely a constant scalar w.r.t. $\bs$.
To deal with the optimization problem of \eqref{eq:mse-aug}, we follow the identical framework as \cite{6472022} that exploits the idea of fractional programming \cite{dinkelbach}. 

Let $a(\bs) = \bs^H\bchi\bs + \mu$, and $b(\bs) = \bs^H\bW\bs > 0$ (MSE needs to be finite). 
Further, let $f(\bs) = a(\bs)/b(\bs)$ and $\bs_*$ denote the current value of $\bs$.
We define $g(\bs)\triangleq a(\bs)-f(\bs_*)b(\bs)$, and $\bs_{\dagger} \triangleq \arg\min_{\bs} g(\bs)$.
It can be easily verified that $g(\bs_{\dagger}) \leq g(\bs_*) = 0$.
As a result, we have that $g(\bs_{\dagger}) = a(\bs_{\dagger})-f(\bs_*)b(\bs_{\dagger}) \leq 0$ which indicates to $f(\bs_{\dagger}) \leq f(\bs_*)$ as $b(\bs_{\dagger}) > 0$. 
Therefore, $\bs_{\dagger}$ can be considered as a new vector $\bs$ that monotonically decreases $f(\bs)$.
Note that $\bs_{\dagger}$ does not necessarily have to be a minimizer of $g(\bs)$; instead, it is enough if $\bs_{\dagger}$ is such that $g(\bs_{\dagger}) \leq g(\bs_*)$.

Under the assumption that $\|\bs\|^2_2=N$, for a fixed $\bw$, and any arbitrary $\bs_*$ of the minimizer $\bs$ of \eqref{eq:opt-s}, we have:
\begin{align}
	g(\bs) = \bs^H(\bchi - f(\bs_*)\bW)\bs + \mu= \bs^H\bT\bs + \mu,
\end{align}
where $\bT \triangleq \bchi - f(\bs_*)\bW$.
Then the problem of \eqref{eq:mse-aug} w.r.t. unimodular $\bs$ can be recast as the following unimodular quadratic program (UQP) \cite{soltanalian2014designing}:
\begin{align}\label{eq:uqp}
	\max_{\bs} ~\bs^H\tilde{\bT}\bs\qquad
	\st ~|s_k|=1, \qquad 1 \leq k \leq N,
\end{align}
where $\tilde{\bT} \triangleq \lambda \bI - \bT$ is a positive definite matrix and $\lambda$ is a real scalar greater than the maximum eigenvalue of $\bT$.
Note that \eqref{eq:uqp} is NP-hard in general, and a sub optimal solution can be sought by semi-definite relaxation (SDR).
To tackle this problem efficiently, in \cite{soltanalian2014designing} a set of \textit{power method-like} iterations was suggested that can be used to monotonically increase the criterion in \eqref{eq:uqp}; namely, the vector $\bs$ is updated in each iteration $n$ using the nearest-vector problem
\begin{align}\label{eq:power}
	\min_{\bs^{(n+1)}} &~{\left\| \bs^{(n+1)} - \tilde{\bT}\bs^{(n)}\right\|_2} \nonumber\\
	\st &~\left|s_k^{(n+1)}\right| = 1, \qquad 1 \leq k \leq N. 
\end{align}
Fortunately, the solution to \eqref{eq:power} is simply given analytically by $\bs^{(n+1)} = e^{j \arg(\tilde{\bT}\bs^{(n)})}$.
A proof of monotonically increasing behavior of the UQP objective in \eqref{eq:uqp} can be found in \cite{6472022}.

\subsection{Optimization of $\bw$}
For a fixed $\bs$, the objective of \eqref{eq:MSEonebit} can further be simplified as,
\begin{align} \label{eq:probI}
\expecE\left\lbrace\frac{\bw^H\bD^{\half} \bar{\bR} \bD^{\half}\bw}{|\bw^H\bs|^2} \right\rbrace&= \frac{\expecE\left\lbrace\tr\left(\bw\bw^H\bD^\half\bar{\bR}\bD^\half\right)\right\rbrace}{|\bw^H\bs|^2} \\
&= \frac{\expecE\left\lbrace\bd^H\left(\bw\bw^H\odot\bar{\bR}^H\right)\bd\right\rbrace}{|\bw^H\bs|^2} \nonumber \\
&= \frac{\tr\left(\left(\bw\bw^H\odot\bar{\bR}^H\right) \expecE\left\lbrace\bd\bd^H\right\rbrace\right)}{|\bw^H\bs|^2} \nonumber.
\end{align}
It is clearly evident that the knowledge of $\bd$ indirectly demands more information about $\beta$ and $\bGamma$.
However, assuming the statistics of the noise is unchanging, one can estimate $\bGamma$ in a normalized sense by just listening to the environment while not transmitting any waveform.
As a result, from the one-bit receiver, the normalized interference covariance matrix $\bar{\bGamma}$ can be obtained in a similar fashion as, $ \bar{\bGamma}  \triangleq \bA^{-\half}\bGamma\bA^{-\half} $, where $\bA = \bGamma \odot \bI$.
Thus the interference covariance matrix $\bR$ in \eqref{eq:R} can be reformulated as,
\begin{align}
\bR = \bD^{\half} \bar{\bR} \bD^{\half} = \beta\bS + \bA^{\half} \bar{\bGamma} \bA^{\half},
\end{align}
where $\bS = \sum_{k\neq 0} {\bJ_k\bs\bs^H\bJ_k^H}$ is constant for a known $\bs$.
Hence, a judicious approach is to solve the following problem in order to optimize $\bd, \ba$, and $\beta$ in a joint manner:
\begin{align}\label{eq:noisy}
\{\hat{\bd}, \hat{\ba}, \hat{\beta}\} = \nonumber\\
\arg\min_{\bd, \ba, \beta} &~~\left\|\Diag(\bd)^{\half}~\bar{\bR}~\Diag(\bd)^{\half}  \right. \nonumber\\
	&\qquad\left. - \beta\bS + \Diag(\ba)^{\half}~\bar{\bGamma}~\Diag(\ba)^{\half}\right\|_F^2, \nonumber\\
\st &~~ \bd > \bzero, \ba > \bzero, \beta>0.
\end{align}
The above minimization problem is non-convex, and hence in order to efficiently solve it, we resort to an alternating approach: by solving for each variable while keeping the other two constant. 
By doing so, w.r.t. each variable the problem becomes convex and can be solved using a number of available numerical solvers, such as the ``fmincon'' function in MATLAB that implements BFGS.
Note that by solving \eqref{eq:noisy}, one can obtain $\beta$ and $\bd$ in an average sense which in other words justifies the usage of expectation in the formulation of \eqref{eq:probI}.\looseness=-1

With this information in mind, let $ \sum_{k=1}^{N}{\nu_k\bu_k\bu_k^H}$ represent the eigenvalue decomposition (EVD) of $\expecE\{\bd\bd^H\}$,
where $\{\nu_k\}$ and $\{\bu_k\}$ are the $k$-th eigenvalue and eigenvector, respectively.
As a result, the numerator of \eqref{eq:probI} can further be simplified as,
\begin{align}
	&\tr\left(\left(\bw\bw^H\odot\bar{\bR}^H\right) \sum_{k=1}^{N}{\nu_k\bu_k\bu_k^H}\right) \nonumber\\
	&\qquad=~~ \sum_{k=1}^{N} {\nu_k \bu_k^H \left(\bw\bw^H\odot\bar{\bR}^H\right)\bu_k} \nonumber\\
	&\qquad=~~ \tr\left(\left(\bw\bw^H\right) \sum_{k=1}^{N} {\nu_k~\Diag(\bu_k)~ \bar{\bR}~\Diag(\bu_k^H)}\right)\nonumber\\
	&\qquad=~~ \bw^H \bQ\bw,
\end{align}
where
\begin{align}\label{eq:Q}
	\bQ = \sum_{k=1}^{N} {\nu_k~\Diag(\bu_k)~ \bar{\bR}~\Diag(\bu_k^H)}.
\end{align}
It is interesting to notice that, $\bQ$ can be viewed as $\expecE\{\bR\}$.
A relevant proof is discussed in Appendix \ref{app:1}.
Finally, the optimization problem translates to,
\begin{align}\label{eq:probER}
\min_{\bw, \bs}~~\frac{\bw^H\bQ\bw}{|\bw^H\bs|^2}.
\end{align}
Hence, for a given $\bs$, the optimization problem in \eqref{eq:probER} w.r.t. $\bw$ results in a closed-from solution: $ \bw = \bQ^{-1} \bs $, within a multiplicative constant. Finally, the algorithm CREW (one-bit) is summarized in Algorithm \ref{alg:screw} in a concise manner.

\section{Numerical Examples and Discussion}
In this section, we evaluate the performance of CREW (one-bit) and compare it with three state-of-the-art methods; namely CAN-MMF, CREW (fre) and CREW (cyclic).
The CAN-MMF method employs the CAN algorithm in \cite{4749273} to simply design a transmit waveform with good correlation properties and independent of the receive filter.
Note that no prior knowledge of interference is used in the waveform design of CAN-MMF.

\begin{algorithm}[t]
	\caption{\textsc{CREW (one-bit)}}\label{alg:screw}
	
	\begin{algorithmic}[1]
		\Ensure $\bs^{(0)} \leftarrow$ unimodular (or low PAR) vector in $\complexC^N$, $\bw^{(0)} \leftarrow$ random vector in $\complexC^N$, the outer loop index $t\leftarrow 1$.
		
		\Repeat
		\State \textbf{For fixed} $\bw$,
		\begin{enumerate}[i:]
			\item Compute $\bchi, \bW$ using \eqref{eq:chi}, and thus, in turn find $\tilde{\bT}$.
			\item Solve the power method like iterations discussed in \eqref{eq:power}, and calculate $\bs^{(t)}$ in each iteration until convergence.
		\end{enumerate}
		
		\State Measure $\bar{\bGamma}$ at the output of the one-bit receiver and compute $\bar{\bR}$ using $\bs^{(t)}$.
		\State \textbf{For fixed} $\bs$,
		\begin{enumerate}[i:]
			\item Solve \eqref{eq:noisy} to obtain $\bd$ and $\beta$ in average sense.
			\item Compute the EVD of $\expecE\{\bd\bd^H\},$ and in turn find $\bQ$.
			\item Update $\bw^{(t)}$ as $\bQ^{-1}\bs^{(t)}$.
		\end{enumerate}
		\Until {convergence, e.g., $\left|\text{MSE}^{(t+1)} - \text{MSE}^{(t)}\right|<\epsilon$ for some given $\epsilon>0$.}
	\end{algorithmic}
\end{algorithm}

We adopt the same simulation setups as in \cite{6472022}.
Especially, for the interference covariance matrix we consider the following:
\begin{align*}
	\bGamma = \sigma_J^2 \bGamma_J + \sigma^2\bI,
\end{align*}
where $\sigma_J^2 = 100$, and $\sigma^2 = 0.1$ are the jamming
and noise powers, respectively.
Furthermore, the jamming covariance matrix $\bGamma_J$ is given by $[\bGamma_J]_{k,l} = \gamma_{k-l}$ where $[\gamma_0, \gamma_1, \cdots, \gamma_{N-1}, \gamma_{-(N-1)}, \cdots, \gamma_{-1}]^T$ can be obtained by an inverse FFT (IFFT) of the jamming power spectrum $\{\eta_p\}$ at frequencies $(p-1)/(2N-1),~p=1,\cdots,2N-1$. 
For CREW(fre) and CREW(cyclic) we fix the average clutter power to $\beta=1$. 
Finally, we use the Golomb sequence in order to initialize the transmit waveform $\bs$ for all algorithms.

We consider two modes of jamming: spot and barrage. Spot jamming is concentrated power directed toward one channel or frequency. 
In our example we use a spot jamming located at a normalized frequency $f_0=0.2$.
On the other hand, barrage jamming is power spread over several frequencies or channels at the same time.
We consider a barrage jamming located in the normalized frequency bands $[f_1,f_2] = [0.2, 0.3]$.

\begin{figure*}
	\centering
	\subfigure[]{	
		\centering
		\includegraphics[draft=false,width=0.7\textwidth]{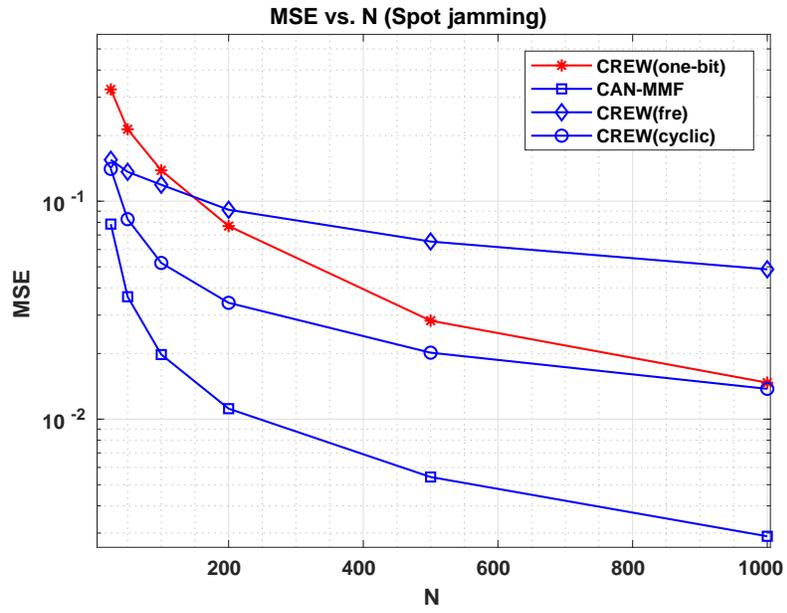}
	}
	\subfigure[]{
		\centering
		\includegraphics[draft=false,width=0.7\textwidth]{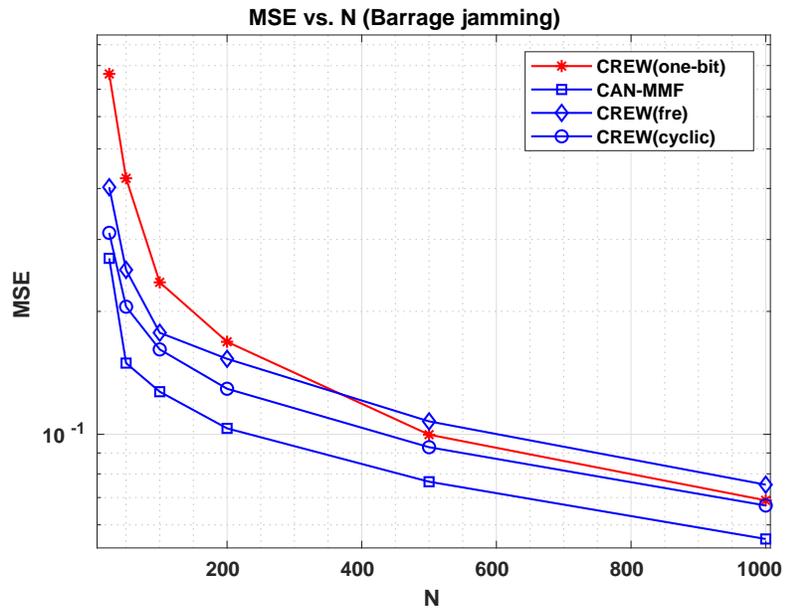}
	}
	\caption{MSE values obtained by the different design algorithms for (a) spot jamming with normalized frequency $f_0=0.2$, and (b) barrage jamming in the normalized frequency interval $[f_1, f_2] = [0.2, 0.3]$ for the unimodularity constraint on the transmit sequence.}
	\label{fig:2}
\end{figure*}

Fig. \ref{fig:2} (a)-(b) depict the MSE values for spot and barrage jamming, respectively, corresponding to CAN-MMF, CREW(fre), and CREW(cyclic), under the unimodularity constraint, for various sequence lengths.
It is evident from the figures that when the sequence length $N$ is small, the MSE is higher for CREW (one-bit) compared to other algorithms.
However, as $N$ increases, CREW (one-bit) shows similar performance as CREW (cyclic) and eventually, they coincide with one another for higher values of $N$.
Consequently, it is implied that higher signal length introduces more degrees of freedom in designing transmit waveform and thus, compensates for the uncertainties in interference statistics.
It is further important to notice that the knowledge of the one-bit measurements impacts the design of the receive filter and alternatively the design of the receive filter coefficients impacts the design of transmit waveform, which justifies the role of a cognitive radar.


\appendices
%

\section{} \label{app:1}
\noindent By using $\bD^{\half} = \Diag(\bd)$, the following can be deduced: 
\begin{align}\label{eq:er}
	\expecE\{\bR\} = \expecE\{\bD^{\half} \bar{\bR} \bD^{\half}\}
	= \expecE\{\bd\bd^H\}\odot\bar{\bR}.
\end{align}
Assuming $\expecE\{\bd\bd^H\} = \bdeta\bdeta^H + \bSigma$, \eqref{eq:er} can reformulated as
\begin{align}
	\expecE\{\bR\} &= (\bdeta\bdeta^H + \bSigma)\odot\bar{\bR} \nonumber \\
	&= \sum_{k=1}^{N}{\nu_k\bu_k\bu_k^H}\odot\bar{\bR} \nonumber \\
	&= \sum_{k=1}^{N} {\nu_k~\diag(\bu_k)~ \bar{\bR}~\diag(\bu_k^H)},
\end{align}
and the proof is complete.


\balance
\bibliographystyle{IEEEbib}
\bibliography{bib-est}

\begin{thebibliography}{10}

\bibitem{6104176}
P.~{Stoica}, H.~{He}, and J.~{Li},
\newblock ``Optimization of the receive filter and transmit sequence for active
  sensing,''
\newblock {\em IEEE Transactions on Signal Processing}, vol. 60, no. 4, pp.
  1730--1740, April 2012.

\bibitem{4644058}
P.~{Stoica}, J.~{Li}, and M.~{Xue},
\newblock ``Transmit codes and receive filters for radar,''
\newblock {\em IEEE Signal Processing Magazine}, vol. 25, no. 6, pp. 94--109,
  November 2008.

\bibitem{6472022}
M.~{Soltanalian}, B.~{Tang}, J.~{Li}, and P.~{Stoica},
\newblock ``Joint design of the receive filter and transmit sequence for active
  sensing,''
\newblock {\em IEEE Signal Processing Letters}, vol. 20, no. 5, pp. 423--426,
  May 2013.

\bibitem{Spafford}
L.~Spafford,
\newblock ``Optimum radar signal processing in clutter,''
\newblock {\em IEEE Trans. Inf. Theor.}, vol. 14, no. 5, pp. 734--743, Sept.
  2006.

\bibitem{8822763}
A.~{Ameri}, A.~{Bose}, J.~{Li}, and M.~{Soltanalian},
\newblock ``One-bit radar processing with time-varying sampling thresholds,''
\newblock {\em IEEE Transactions on Signal Processing}, vol. 67, no. 20, pp.
  5297--5308, Oct 2019.

\bibitem{8683876}
S.~{Khobahi}, N.~{Naimipour}, M.~{Soltanalian}, and Y.~C. {Eldar},
\newblock ``Deep signal recovery with one-bit quantization,''
\newblock in {\em ICASSP 2019 - 2019 IEEE International Conference on
  Acoustics, Speech and Signal Processing (ICASSP)}, May 2019, pp. 2987--2991.

\bibitem{5955138}
J.~N. {Laska}, Z.~{Wen}, W.~{Yin}, and R.~G. {Baraniuk},
\newblock ``Trust, but verify: Fast and accurate signal recovery from 1-bit
  compressive measurements,''
\newblock {\em IEEE Transactions on Signal Processing}, vol. 59, no. 11, pp.
  5289--5301, Nov 2011.

\bibitem{2019arXiv191007591B}
A.~{Bose}, S.~{Khobahi}, and M.~{Soltanalian},
\newblock ``{Joint Optimization of Waveform Covariance Matrix and Antenna
  Selection for MIMO Radar},''
\newblock {\em arXiv e-prints}, p. arXiv:1910.07591, Oct 2019.

\bibitem{8645383}
S.~{Khobahi} and M.~{Soltanalian},
\newblock ``Signal recovery from 1-bit quantized noisy samples via adaptive
  thresholding,''
\newblock in {\em 2018 52nd Asilomar Conference on Signals, Systems, and
  Computers}, Oct 2018, pp. 1757--1761.

\bibitem{van1966spectrum}
J.~H. V.~Vleck and D.~Middleton,
\newblock ``The spectrum of clipped noise,''
\newblock {\em Proceedings of the IEEE}, vol. 54, no. 1, pp. 2--19, 1966.

\bibitem{bussgang1952crosscorrelation}
J.~J. Bussgang,
\newblock ``Crosscorrelation functions of amplitude-distorted {Gaussian}
  signals,''
\newblock 1952.

\bibitem{liu2017one}
C.-L. Liu and P.~P. Vaidyanathan,
\newblock ``One-bit sparse array {DOA} estimation,''
\newblock in {\em Acoustics, Speech and Signal Processing (ICASSP), 2017 IEEE
  International Conference on}. IEEE, 2017, pp. 3126--3130.

\bibitem{dinkelbach}
W.~Dinkelbach,
\newblock ``On nonlinear fractional programming,''
\newblock {\em Management Science}, vol. 13, no. 7, pp. 492--498, 1967.

\bibitem{soltanalian2014designing}
M.~Soltanalian and P.~Stoica,
\newblock ``Designing unimodular codes via quadratic optimization,''
\newblock {\em IEEE Transactions on Signal Processing}, vol. 62, no. 5, pp.
  1221--1234, 2014.

\bibitem{4749273}
P.~{Stoica}, H.~{He}, and J.~{Li},
\newblock ``New algorithms for designing unimodular sequences with good
  correlation properties,''
\newblock {\em IEEE Transactions on Signal Processing}, vol. 57, no. 4, pp.
  1415--1425, April 2009.

\end{thebibliography}
\end{document}